\documentclass[pre,aps,showpacs,twocolumn]{revtex4}
\usepackage[T1]{fontenc}
\usepackage[utf8]{inputenc}
\usepackage[french]{babel}
\usepackage{amsmath}
\usepackage{amssymb}
\usepackage{epsfig}
\usepackage{amscd}
\usepackage{verbatim}
\usepackage{color}

\usepackage{pstricks}
\usepackage{psfrag}

\newcommand{\elabel}[1]{\label{e:#1}}

\newcommand{\eq}[1]{Eq.~(\ref{e:#1})}

\newcommand{\fig}[1]{Fig.~\ref{f:#1}}

\newcommand{\quot}[1]{``#1''}

\newcommand{\wfigure}[3][\columnwidth]{
\begin{figure}[ht]
  \centerline{\includegraphics[width=#1]{#2}}
  \caption{#3}
  \label{f:#2}
\end{figure}}

\begin{document}


\title{Event-driven Monte Carlo algorithm for general potentials}
\author{Etienne P. Bernard}
\email{etienne.bernard@mit.edu}
\affiliation{Department of Physics, Massachusetts Institute of Technology\\
Cambridge, Massachusetts 02139, USA}
\author{Werner Krauth}
\email{werner.krauth@ens.fr}
\affiliation{ Laboratoire de Physique Statistique, Ecole Normale
Supérieure, UPMC, CNRS\\ 24 rue Lhomond, 75231 Paris Cedex 05, France}

\begin{abstract}
We extend the event-chain Monte Carlo algorithm from hard-sphere
interactions to the micro-canonical ensemble (constant potential
energy) for general potentials. This event-driven Monte Carlo
algorithm is non-local, rejection-free, and allows for the breaking of
detailed balance. The algorithm uses a discretized potential, but its
running speed is asymptotically independent of the discretization.  We
implement the algorithm for the cut-off linear potential, and discuss
its possible implementation directly in the continuum limit.
\end{abstract}
\maketitle

The event-chain algorithm \cite{Bernard_2009}, a non-local, rejection-free
Markov-chain algorithm for hard-sphere systems, has proved considerably
faster than the local Monte Carlo algorithm.  It has allowed us to show
that two-dimensional melting in hard disks proceeds via a first-order
liquid-hexatic transition \cite{Bernard_2011}.

In the present paper, we generalize the event-chain algorithm to the
case of general potentials, greatly extending the scope of the
original method. As for event-driven molecular dynamics (MD), we
discretize the potential (here on an energy scale $\Delta_E$), but
unlike the MD algorithm \cite{Event_driven_MD,Bannerman}, simulations
at arbitrary small $\Delta_E$ are feasible.

\begin{figure}[ht!]
   \psfrag{l}[cc][cc]{\large{$\ell$}}
   \center
   \includegraphics[width=.33\textwidth]{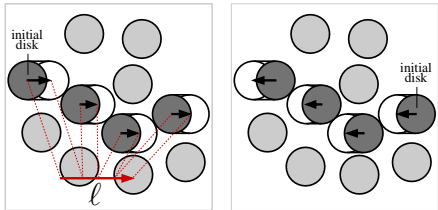}
   \caption{ Event-chain algorithm for hard disks
   \cite{Bernard_2009}. \emph{Left:} Collective move, with a total
   displacement $\ell$ as a fixed parameter of the algorithm. 
   The move is never rejected.  \emph{Right:}
   Return move, demonstrating microreversibility.}
\label{f:event_chain_move}
\end{figure}


In the event-chain algorithm \cite{Bernard_2009}, a randomly chosen disk
is moved in a straight line by a distance $\ell$, if this generates
no overlap for all $\ell' \le \ell$, or else until it hits another
particle $i'$. The latter particle, $i'$, then moves for the remainder
of the distance $\ell$ or, again, until it hits yet another particle,
etc. (see \fig{event_chain_move}). The event-chain move is micro-reversible, and
it satisfies detailed balance if all the displacements are in the same
direction. For hard disks, it is about two orders of magnitude faster
than the local Monte Carlo algorithm \cite{Metropolis}, and it compares
favorably with the MD method\cite{Alder_Wainwright}.


In the event-driven Monte Carlo algorithm, we consider a discretized
potential $V(r)$, with discontinuities such that $V(r)$ and thus the
total potential energy are multiples of a given energy step
$\Delta_E$ \cite{footnote}.
From an initial configuration of energy $E$, a
randomly chosen particle $i$ is moved in a straight line until the
total displacement, \emph{at energy $E$ only}, equals a fixed value
$\ell$ or else until it \quot{hits} another particle $i'$. This means
that the energy of the system would exceed $E$ if the particle $i$ moved
any farther.  The latter particle, $i'$, then moves for the remainder
of the distance $\ell$ or, again, until it hits yet another particle,
etc. (see \fig{Event_stepped_move}).  In the absence of boundary
effects, the move is guaranteed to terminate. It is again
rejection-free and micro-reversible.  Detailed balance may again be
broken and the algorithm can be run with moves, say, in the $+x$ and
$+y$ direction only.  
\wfigure[6cm]{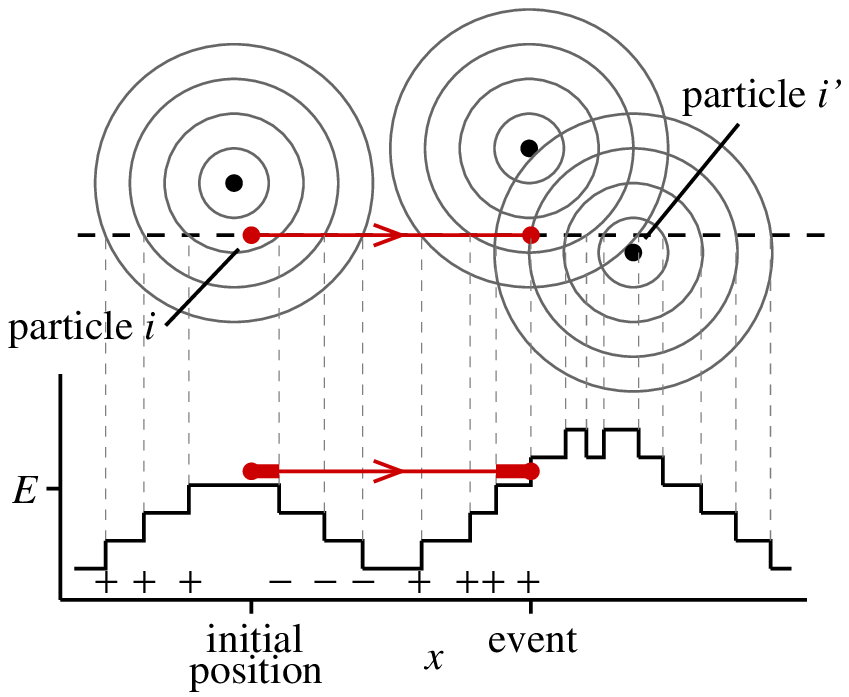}{Event-driven
  Monte Carlo displacement for $N=4$. Particle $i$ moves from its
  initial configuration until its collision event with particle $i'$
  (potential of \eq{stepped_potential} with $\Delta_E = 1/4$ and
  $E=3/4$). Only displacements at energy $E$ are discounted from
  $\ell$.}

The event-driven Monte Carlo algorithm can be implemented by computing
the intersection points of the potential discontinuities with the
trajectory of the particle $i$.  One recovers $E(x)$ by adding up the
signs in the sorted list of the intersections.  This is illustrated in
\fig{Event_stepped_move} for the potential defined in
\eq{stepped_potential}. In \fig{gr_local_ec}, we consider particles
interacting with the potential
\begin{equation}
V(r)= \begin{cases}
       0 & \text{if $r>1$}\\
       1 - \Delta_E \lfloor r/\Delta_E \rfloor & \text{otherwise}
      \end{cases}
\elabel{stepped_potential}
\end{equation}
for $\Delta_E=1/10$. We compare the pair-correlation function of the
event-driven Monte Carlo algorithm for a large system with $N=128^2$
particles to the results obtained with a local Monte Carlo algorithm
in the same system. With the local Monte Carlo algorithm, only
configurations with potential energy $E$ are used. There is no doubt
about the correctness of the algorithm and the implementation. With this
simple implementation, we reached $\sim 5 \times 10^{9}$ collisions per
hour on a 3 GHz workstation.

\wfigure[5cm]{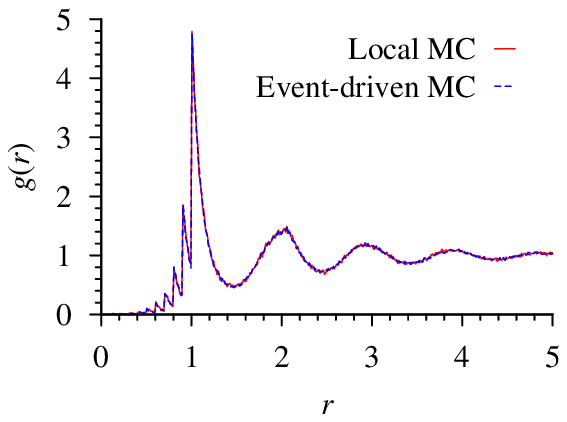}{Pair-correlation function computed with the
  local and the event-driven Monte Carlo algorithms ($N=128^2$, box of
  size $128^2$, $E = N\Delta_E $, $\Delta_E=0.1$). Data agree, which
  confirms the correctness of the algorithm.}


In the naive implementation suggested in \fig{Event_stepped_move}, the
number of steps between events scales $\propto 1/\Delta_E$, and the
algorithm becomes very slow in the limit $\Delta_E \to 0$. Unlike for
event-driven molecular dynamics \cite{Event_driven_MD}, this
difficulty can be overcome, and the speed of the algorithm remains
constant for $\Delta_E \to 0$. The algorithm is straightforward for a
convex potential with finite support, as for example the
continuum limit of \eq{stepped_potential}, given by
\begin{equation*}
V_{\text{cont}}(r)= \begin{cases}
       0         & \text{if $r>1$}\\
       1 - r     & \text{otherwise}
      \end{cases},
\elabel{continuous_cropped_potential}
\end{equation*}
for which $E(x)$ is continuous and piecewise $C^\infty$ while
$\partial E/ \partial x$ is piecewise monotonously decreasing (see
\fig{Continuous}).  The root $E_{\text{cont}}(x_{\text{root}})=E$ of the
continuous potential is uniquely determined \emph{via} a decision tree
in subsequent $C^{\infty}$ intervals of $E(x)$. For finite $\Delta_E$,
one must only identify the discontinuity steps of the potential in the
interval $x_{\text{root}} \pm \text{const} [\partial E / \partial x
(x_{\text{root}})]^{-1}$. We have implemented the algorithm for very
large values of $1/\Delta_E$ (with $\ell \propto \Delta_E$), and achieved
constant scaling of the algorithm.

\wfigure[7cm]{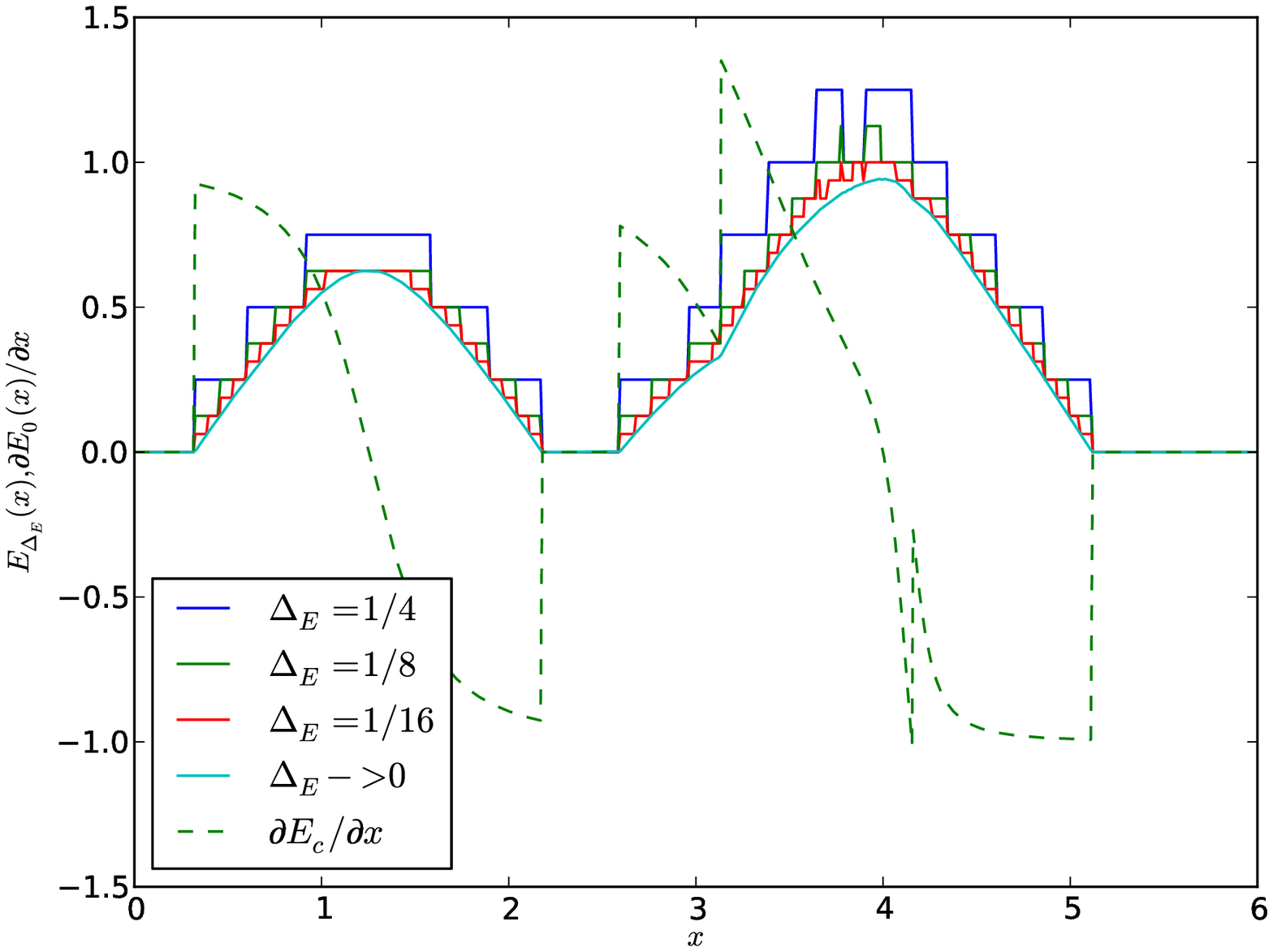}{Energy $E(x)$ for the configuration of
\fig{Event_stepped_move}, for different discretizations, and in the
continuum limit, and derivative $\partial E_{\text{cont}}/\partial x$
which is piecewise monotonously decreasing.}


Although the event-driven Monte Carlo algorithm can be implemented
for arbitrarily small $\Delta_E$, and does not slow down then, it is
interesting to consider the $\Delta_E \to 0$ limit.  A good model for this
is to suppose that rather than being fixed, $\Delta_E^{-1}$ is sampled,
between events, from a distribution with very large mean. The collision
rules in \fig{Event_stepped_move} then become probabilistic. They are
readily analyzed in simple cases. The event itself, which takes place
at energy $E$, is initiated and ended  by particles coming in from and
moving out to lower energies.  This event may involve a sizable number of
particles and thus become quite complex. It is unclear whether truly
general stochastic rules for the collisions and the use of  $\ell$
can be obtained.


In conclusion, in this paper, we have generalized the successful
event-chain algorithm from hard spheres to the case of general
potentials and provided details on its implementation. This algorithm
is an alternative to the local Monte Carlo algorithm \cite{Metropolis}
and, even with a simple implementation, should outperform it in most
applications. Moreover, arbitrarily small discretization steps can be
handled efficiently, which is not the case in the event-driven MD
\cite{Alder_Wainwright,Event_driven_MD,Bannerman}.


\begin{thebibliography}{99}
\bibitem{Bernard_2009} E. P. Bernard, W. Krauth, D. B. Wilson,
\emph{Phys. Rev. E} 80 056704 (2009).

\bibitem{Bernard_2011} E. P. Bernard, W. Krauth,
\emph{Phys. Rev. Lett.} 107, 155704 (2011).

\bibitem{Event_driven_MD} G. A. Chapela, L. E. Scriven, and H. T. Davis,
\emph{J. Chem. Phys.} 91, 4307 (1989)

\bibitem{Bannerman} M. N. Bannerman, R. Sargant, and L. Lue 
  \emph{J. Comput. Chem.} 32, 3329 (2011)

\bibitem{Metropolis} N. Metropolis, A. W.  Rosenbluth, M. N. Rosenbluth,
A. H. Teller, E.  Teller, \emph{J. Chem. Phys.} 21, 1087 (1953).

\bibitem{Alder_Wainwright} B. J.  Alder  and T. E. Wainwright,
\emph{J. Chem. Phys.} 31, 459 (1959)

\bibitem{footnote}
$V(r)$ can be discretized by energy step $\Delta_E$, or multiples of
$\Delta_E$ for strongly varying potentials.

\end{thebibliography}
\end{document}